\documentclass[12pt,english,floatfix,superscriptaddress,aps,prd,preprint,showkeys,nofootinbib]{revtex4}
\usepackage{amsmath}
\usepackage{amssymb}
\usepackage{amsbsy}
\usepackage{amsfonts}
\usepackage{amsopn}
\usepackage{amstext}
\usepackage{graphicx}
\usepackage{amssymb}
\usepackage{amsfonts}
\usepackage{amsmath}
\usepackage{graphicx}
\usepackage[english]{babel}
\usepackage{color}
\usepackage{slashed}
\usepackage{esint}
\usepackage[dvips]{epsfig}
\usepackage[dvips]{graphicx}
\usepackage{float}
\usepackage{units}
\usepackage{textcomp}

\usepackage{hyperref}
\usepackage{slashed}

\begin{document}

\title{Symmetries and fields in Randers-Finsler spacetime}


\author{J. E. G. Silva}
\email{silvajeg@gmail.com}
\affiliation{Indiana University Center for Spacetime Symmetries (IUCSS), Bloomington, Indiana 47405, USA}





\date{\today}

\begin{abstract}
We study the symmetries of the Lorentz violating Randers-Finsler spacetime.
The privileged frame defined by the background vector diagonalises the deformed mass-shell and provides an anisotropic observer transformations  
The particle Randers transformations
are achieved through the Finslerian Killing equation which also shows the equivalence between observer and particle Randers transformation. For a constant background vector, the Randers transformations can be regarded as a deformed directional-dependent Poincar\'{e} transformation.
The Randers algebra is a deformed Poincar\'{e} algebra whose structure coefficients are given by the Randers metric.
We propose a field theory invariant over the Randers transformations. The dynamics exhibits nonlocal operators which yields to Lorentz violating
terms of the nonminimal SME.
The relation and implications of the Randers algebra with other Lorentz violating theories are discussed.
\end{abstract}




\maketitle

\section{Introduction}
At Planck scale, several theories including string theory \cite{KS}, noncommutative geometry \cite{noncommutativegeometry}, Horava gravity \cite{horava}, and Very Special Relativity \cite{vsr} suggest that some low-energy symmetries, as the CPT and Lorentz symmetry, may be violated. A theoretical framework to study the effects of reminiscent Lorentz-CPT breaking at intermediate scales is the so-called Standard Model Extension (SME) \cite{cptviolation}. Lorentz violating effects studies have been carry out in a broad class of phenomena, ranging from muons, neutrinos, photons, hydrogen atom, among other \cite{smeapplications}. For a comprehensive review of Lorentz-violation tests, see Ref.\cite{rmp}.

The study of Lorentz violation on gravity is more complex. In flat spacetimes, the Lorentz violation can be driven by background constant tensors (explicit Lorentz violation) that break the particle Lorentz symmetry but keep the observer Lorentz symmetry. Nevertheless, in curved spacetimes, the explicit Lorentz violation is incompatible with the Biachi identities \cite{kosteleckysmegravity}. One way to overcome this is assuming the Lorentz violating tensors are created by the vacuum expectation value of a vector field, as the bumblebee \cite{bumblebee} or the aether fields \cite{Jacobson:2000xp} undergone a spontaneous symmetry process \cite{kosteleckysmegravity}. Another possibility is through an anisotropic geometry, the so-called Finsler geometry.

In Finsler geometry, the length of vector is measured with a general norm, called Finsler function \cite{chern,rund,miron}. The Riemann geometry is a special case for which the Finsler function is quadratic on the vector components. Since the background tensor modifies the dispersion relation of the fields, the point particle classical Lagrangians and mass shell can be described by a class of Finsler functions, called SME-based Finsler geometries \cite{Kosteleckyclassicallagrangians,Kosteleckyriemannfinsler}. Each Lorentz-violating coefficient provides a different SME-based Finsler structure, as the Randers \cite{Kosteleckyriemannfinsler}, b-space \cite{Kosteleckyriemannfinsler,foster}, bipartite \cite{Kosteleckybipartite,euclides}  and other spacetimes \cite{ColladayMc,russell,marco}. Other Lorentz violating theories also has Finsler extensions for curved spacetimes, as the DSR \cite{dsrfinsler} and VSR \cite{Gibbons}, whose the curved extension is the Bogoslovsky spacetime \cite{bogoslovsky}.

A special Finsler structure is given by the Rander spacetime, where the anisotropy is driven by a background vector \cite{randers,beem,bejancu}. The Randers interval has the usual quadratic Lorentz interval added with a perturbative linear projection of the 4-velocity into the background vector \cite{randers}. In the context of the SME, the Randers spacetime can be regarded as a classical description of a fermion subjected to a CPT-Odd Lorentz-violating coefficient \cite{Kosteleckyriemannfinsler}. The particle dynamics in Randers spacetime is analogous to the Lorentzian invariant dynamics under the influence of a background electromagnetic vector field $a^\mu$. The effects of the Rander spacetime anisotropy have been studied in cosmology \cite{randerscosmology} and astrophysics \cite{randersastrophysics}.

In this work we study the symmetry transformations of the Randers interval, and hence, of the particle action. We extend the previous analysis performed in Ref.\cite{girelli,changmdr}. By employing a diagonalization of the Randers mass-shell, we show that the Randers observer transformations can be realized as the product of the Lorentz transformation with a deformation of the mass shell provided by the eigenvalues of the mass shell and a boost due to the privileged frame. Using the Killing equation and the vielbein formalism we obtain the Randers transformations. The background vector modifies the generators of the Randers algebra where the Finsler metric can be seen as the structure coefficients of a deformed Poincar\'{e} algebra.

Using Randers-Finsler invariant objects and assuming the momentum-dependence of the Finsler metric can be regarded as an operator on fields, we propose
an anisotropic dynamics for the scalar, vector, fermionic and gravitational fields. This metric operator approach shares some resemblance with the noncanonical kinetic models \cite{noncanonicalkinetic} and enable us to define a field theory on the spacetime and not on the tangent bundle \cite{vacaru,pfeifer,Perlick,Lammerzahl}. For Randers spacetime this instrinsic dynamics exhibits nonlocal operators, as those found in VSR \cite{vsr} and in Bogoslovksy spacetime \cite{bogoslovsky}. The perturbative character of the Randers anisotropy allow us to expand the nonlocal operators what yields to Lorentz violating terms as found in the SME \cite{cptviolation} and in the Carroll-Field-Jackiw model  higher derivative terms as found in the nonminimal SME \cite{gaugesmenonminimal,fermionsmenonminimal,gravitynomminimalsme}.

This work is organized as the following. In section \ref{section1} we present the definition and main properties of the Randers-Finsler spacetime, as well as the deformed mass shell. In the section \ref{section2} we study the group of coordinate observer transformations which preserve the deformed Randers mass-shell. 
In section \ref{section3} we analyse the particle transformation by means of the Finslerian Killing equation. 
In section \ref{section4} we study the algebra of the generators of the Randers transformations, showing that the Finsler metric acts as deformed Poincar\'{e} algebra structure constant. In section \ref{section5} we propose e study the dynamics of fields on the Randers spacetime. Final comments and relations of the Randers algebra to other Lorentz violating and deformed algebras models are outlined in section \ref{section6}.


\section{Randers-Finsler spacetime}
\label{section1}

In this section we review the definition and the main properties of the Randers-Finsler spacetime. We show how the background vector is included into the geometric structure of the spacetime and this yields to modification of the particle dynamics and mass shell.

In Randers-Finsler spacetime, the infinitesimal intervals of two nearby events $x^\mu,x^\mu +\dot{x}^\mu dt$ are measured by a modified norm, called the Finsler function,  $ds_R:=F(x,\dot{x})dt$, where \footnote{We adopt the mostly plus convention $(-,+,+,+)$ for the metric signature.} \cite{Kosteleckyriemannfinsler,randers}
\begin{eqnarray}
ds_{R}	&	:=	&	\sqrt{-g_{\mu\nu}(x)\dot{x}^{\mu}\dot{x}^{\nu}} + \zeta a_{\mu}(x)\dot{x}^{\mu}.
\end{eqnarray}
The Randers-Finsler has the Lorentz invariant interval $\alpha(x,\dot{x}):=\sqrt{-g_{\mu\nu}(x)\dot{x}^{\mu}\dot{x}^{\nu}}$ and a linear term $\beta(x,\dot{x}):=\zeta a_{\mu}(x)\dot{x}^{\mu}$ which drives the Lorentz violation. The presence of the small parameter $0\leq \zeta <1$ accounts for the perturbative character of the Lorentz violation \footnote{Assuming $\zeta=\frac{M}{M_{Pl}}$, where $M_{Pl}$ is the Planck scale, the linear term $\beta$ describes an effective anisotropic geometry due to reminiscent quantum effects for energies $M<M_{Pl}$.} \cite{Kosteleckyriemannfinsler}.

The modified interval can be rewritten in terms of a anisotropic or Finsler metric, by $ds_{R}	=	\sqrt{-g_{\mu\nu}^F(x,\dot{x})\dot{x}^{\mu}\dot{x}^{\nu}}dt$ \cite{beem,bejancu}. Thus, even for a flat background metric $g_{\mu\nu}=\eta_{\mu\nu}$, a varying background vector $a^\mu$ provides analogue noninertial effects. In the Randers-Finsler spacetime, the Finsler metric has the form \cite{chern}
\begin{equation}
\label{randersmetric}
g_{\mu\nu}^{F}(x,\dot{x})=\frac{F}{\alpha}g_{\mu\nu} - \frac{\beta}{\alpha}U_{\mu}U_{\nu} + 2\zeta  U_{(\mu}a_{\nu)} + \zeta^{2}a_{\mu}a_{\nu},
\end{equation}
where $U_{\mu}:=g_{\mu\rho}U^\rho$ and $U^\rho:=\frac{\dot{x}^\rho}{\alpha(\dot{x})}$ is the Lorentzian unit 4-velocity.

The Randers-Finsler interval $ds_R$ provides a modified action for a massive point particle, as \cite{randers}
\begin{equation}
\label{finsleraction}
S^{F}	:=	-m\int_{I}{\left(\sqrt{-g_{\mu\nu}\dot{x}^{\mu}\dot{x}^{\nu}} + \zeta a_{\mu}(x)\dot{x}^{\mu}\right)dt}.
\end{equation}
The parametrization invariance of the action provides two important consequences. First, the background vector is gauge invariant, $a'_{\mu}=a_\mu + \partial_\mu \Phi$ \cite{randers}. Second, the particle proper time is anisotropic, i.e., $d\tau_F:=F(x,\dot{x})dt=\gamma^{-1}_{R}(v,a)dt$, where $\gamma_{R}(v,a):=\frac{1}{\sqrt{1-v^2}+\zeta(a_0 + (\vec{a}\cdot \vec{v}))}$.

From the action \eqref{finsleraction} we can obtain the Randers-Finsler canonical momentum covector as \cite{Kosteleckyriemannfinsler,changmdr}
\begin{equation}
\label{randersmomentumcovariant}
P_{\mu}^{R}	:=	\frac{\partial L^{R}}{\partial \dot{x}^{\mu}}=P_{\mu} - m\zeta a_{\mu},
\end{equation}
where $P_{\mu}:=mg_{\mu\nu}U^{\nu}$. The Randers-Finsler canonical momentum covector can be rewritten as $P_{\mu}^R=m\omega_{\mu}$, where the unit covector $\omega:=\frac{\partial F}{\partial \dot{x}^{\mu}} dx^{\mu}$ is the so-called Hilbert 1-form \cite{chern}.

Since the action, is defined through the Randers-Finsler metric $g^F(x,\dot{x})_{\mu\nu}$, the relation between the canonical momentum vector $P^{R\mu}$ and the canonical momentum covector $P^R_\mu$ is given by $P^R_\mu:=g^F_{\mu\nu}(x,\dot{x})P^{R\nu}$.
The vector satisfying this anisotropic duality is \cite{girelli}
\begin{equation}
\label{randersmometumvector}
P^{R\mu}:=m\frac{\dot{x}^{\mu}}{F(x,\dot{x})}.
\end{equation}
The Randers-Finsler momentum vector $P^{R\mu}$ can be rewritten as $P^{R\mu}=mU^{R\mu}$, where the $U^{R\mu}:=\frac{dx^\mu}{d\tau_F}=\frac{\dot{x}^{\mu}}{F(x,\dot{x})}$ is the Finslerian unit 4-velocity \footnote{In the mathematical literature, this vector is called Distinguished or Liouville section and it is the Finslerian dual of the Hilbert form $\omega$  \cite{chern}.} \cite{chern}.
The Randers and the Lorentz canonical momentum vector are related by the anisotropic factor $P^{F\mu}=\frac{\alpha (P)}{F(x,P)}P^{\mu}$.

The modified mass shell in the Finsler spacetimes is taken using the Finsler metric, i.e., \cite{bogoslovsky,girelli,vacaru,pfeifer,Perlick,Lammerzahl}
\begin{equation}
g^{F}_{\mu\nu}(x,P^F)P^{F\mu}P^{F\nu}=-m^2.
\end{equation}
The dependence of the Finsler metric on the momentum yields to nonquadratic dispersion relations. The modified mass shell
can be rewritten as $F^2(x,P^R)=m^2$. For the Randers-Finsler spacetime, the modified mass shell has the form
\begin{equation}
\label{randersmassshell}
(g_{\mu\nu}+\zeta^2 a_{\mu}a_{\nu})P^{R\mu}P^{R\nu} - 2\zeta m a_{\mu}P^{R\mu}=-m^2.
\end{equation}


\section{Randers transformations}
\label{section2}

In this section we study the transformations which leaves invariant the modified Randers-Finsler mass shell \eqref{randersmassshell}. Firstly, let us distinguish between two classes of such transformations. The observers transformations are those which the particle is kept fixed and the coordinate system is changed \cite{cptviolation,kosteleckysmegravity}. The second class is the particle transformation where the coordinate system and the background vector are held fixed and an active transformation is acted upon the particle.

\subsection{Observer transformations}

In Randers-Finsler spacetime, the presence of the background metric $\eta_{\mu\nu}$ and the background covector $a_{\mu}$, which are Lorentz invariant, turns the Randers interval $ds_R$ also Lorentz-invariant by observer transformation, i.e., $x'^\mu=\Lambda^{\mu}_{\nu}x^{\nu}$ .
This invariance can be physically interpreted as being the result of measurement of observers whose instruments (rules and clocks) are insensitive to the Lorentz violation effects.

Another set of symmetry transformations can be obtained by considering the rigid motion of the deformed mass shell immersed into a background Lorentz-invariant space. Defining the symmetric tensor $s_{\mu\nu}:=\eta_{\mu\nu}+\zeta^2 a_\mu a_\nu$,
the mass shell is given by the quadric $s_{\mu\nu}P^{R\mu}P^{R\nu} -2m\zeta a_{\mu}P^{R\mu} = -m^2$. Since the mass shell represents the Randers-Finsler norm of the canonical momentum, $F_R (x,P^R)=m$, the general causal surface for a general event $X^\mu$ with norm $F(X)=k$ is given by
\begin{equation}
s_{\mu\nu}X^{\mu}X^{\nu} -2k\zeta a_{\mu}X^{\mu} = -k^2.
\end{equation}

In Randers-Finsler spacetime, the tensor $s_{\mu\nu}$ deforms the quadric by introducing a privileged frame that diagonalizes
$s_{\mu\nu}$. It turns out that $a^\mu$ is an eigenvector of the self-adjoint operator $s^\mu _\nu:=\delta^\mu_\nu + \zeta^2 a^\mu a_\nu$, i.e., $s^\mu_\nu a^\nu= \lambda_a a^\mu$, whose corresponding eigenvalue is $\lambda_a=1+\zeta^2(a_\mu a^\mu)$.
If $a^\mu$ is a Lorentzian timelike vector, then the other eigenvectors are spacelike vectors orthogonal to $a^\mu$ and with eigenvalues equals to one. Then, the eigenvectors of $s^\mu_\nu$ form a privileged frame $B_p=\{\frac{a^\mu}{a},\hat{e}_{1},\hat{e}_{2},\hat{e}_{3}\}$. If $a^\mu$ is a spacelike vector, then the basis of eigenvector $B_p=\{\hat{e}_0,\frac{a^\mu}{a},\hat{e}_{2},\hat{e}_{3}\}$ form a privileged symmetry frame.

Since $0\leq \zeta<1$, the $s_{\mu\nu}$ has still one negative and three positive eigenvalues. At the privileged frame, if $a^\mu$ is timelike the mass shell is shrunk in the direction of $a^\mu$ and left unmodified in the spatial directions; if $a^\mu$ is spacelike, then the mass shell is an elliptic two-sheet hyperboloid whose constant time sections are ellipsoids and the major axis is stretched at the direction of $a^\mu$.

Suppose that in some observer frame $B=\{\hat{e}_0,\hat{e}_1,\hat{e}_2,\hat{e}_3\}$, the timelike background vector field has components $a^\mu=(a^0,a^1,a^2,a^3)$.
Then in the privileged frame $B_p$, the coordinate changes as $\tilde{X}^\mu=(P^{-1})^\mu_\nu X^{\nu}$, where $P^\mu_\nu $
is the change of basis operator from $B$ to $B_p$. If we already start at the privileged frame, this transformation is naturally the identity. The linear term $-2\zeta a_{\mu}X^{\mu}$ can be absorbed into a new coordinate change by $\hat{X}^\mu:=M^\mu_\nu \tilde{X}^\nu+b^\mu$, where $M_{\mu\nu}:=Diag(-\sqrt{\lambda_a},1,1,1)$ and
$b^\mu:=-\zeta\frac{a^\mu}{\lambda}$. The matrix $M_{\mu\nu}$ formed with the squared roots of the eigenvalues of $s_{\mu\nu}$ measures the deformation of the mass shell in different directions.

Applying the transformations $P^{-1}$ and $M$, the mass shell quadric turns into $\eta_{\rho\sigma}\hat{X}^\rho \hat{X}^{\sigma}+R^2=0$, where $R^{2}:=k^2+\frac{\zeta^2 a^2}{1+\zeta^2 a^2}$. Hence, for a timelike or spacelike $a^\mu$, the radius of the elliptic hyperboloid is changed but for a lightlike $a^\mu$ the radius is unchanged.

After the mass shell be completely quadratic, we can apply a Lorentz transformation. Therefore, the symmetry transformations of the mass shell in the Randers-Finsler spacetime has the form
\begin{eqnarray}
\label{randersobservertransformation}
X'^\mu &	=	&	(\Lambda_{(RO)})^\mu_\nu (a)X^\nu + b^{\mu},
\end{eqnarray}
where the Randers-observer transformation is given by
\begin{eqnarray}
(\Lambda_{(RO)})^\mu_\nu (a) &	=	&	\Lambda^\mu_\rho M^\rho_\sigma (P^{-1})^\sigma_\nu,
\end{eqnarray}
and the vector $b^\mu$ is given by $b^\mu:=-\zeta\frac{a^\mu}{\lambda}$.

The anisotropic Randers observer transformation in Eq.\eqref{randersobservertransformation} is similar to one found in \cite{changmdr}. The Randers observer takes into account the anisotropy by considering the existence of the privileged frame
and the deformation of the mass shell, given respectively by the transformations  $(P^{-1})^\sigma_\nu$ and $M^\rho_\sigma$.
Since the transformation $(P^{-1})^\sigma_\nu$ is orthogonal, it can be regarded as a sort of boost between the initial and the privileged frame.

The Randers observer transformation has determinant $\det(\Lambda_{(RO)}(a))=\det M(a) = \sqrt{1+\zeta^2 a^2}$,
hence, this transformation is not orthogonal with respect to the background metric $\eta_{\mu\nu}$. This is an expected result, for the Randers transformation preserves the Randers metric not the Minkowsky metric. Further, since $\sqrt{g^F_{\mu\nu}(x,a)}=(1+\zeta^2 a^2)^{\frac{5}{2}}$, the Randers-Finsler invariant volume is modified by the background vector $a_\mu$.

Since the background metric $\eta_{\mu\nu}$ and vector $a_\mu$ are covariant tensors, they change under the Randers- observer coordinate transformation \eqref{randersobservertransformation} as $a'_\mu =(\Lambda_{RO}^{-1})^\nu_\mu a_\nu $
and $\eta'_{\mu\nu}=(\Lambda_{RO}^{-1})_\mu^{\rho}(\Lambda_{RO}^{-1})_\nu^{\sigma}\eta_{\rho\sigma}$
The induced transformations provides
$\alpha'(\dot{X}'):=\sqrt{\eta'_{\mu\nu}\dot{X}'^{\mu}\dot{X}'^{\nu}}=\sqrt{\eta_{\rho\sigma}\dot{X}^{\rho}\dot{X}^{\sigma}}=\alpha(\dot{X})$ and $\beta'(\dot{X}'):=\zeta a'_{\mu}\dot{X}'^{\mu}=\beta(\dot{X})$ and hence, $F'(\dot{X}')=F(\dot{X})$.
Therefore, $g'^F_{\mu\nu}(\dot{X'})\dot{X}'^{\mu}\dot{X}'^{\nu}=g^F_{\rho\sigma}(\dot{X})\dot{X}^{\rho}\dot{X}^{\sigma}$, what yields to
\begin{equation}
g'^F_{\mu\nu}(X',\dot{X'})(\Lambda_{RO})^{\mu}_\rho (\Lambda_{RO})^{\nu}_\sigma=g^F_{\rho\sigma}(X,\dot{X}).
\end{equation}
Accordingly, the Randers-observer transformation is orthogonal with respect to the Randers-Finsler metric.

The transformations $(\Lambda_{RO})^{\mu}_\rho$ forms the symmetry group of the Randers-Finsler spacetime under observer coordinate transformations. Consider a 1-parameter group of symmetries that $g'^F_{\mu\nu}(X',\dot{X'}) (\Lambda_{RO})^{\mu}_\rho (\lambda)|_{\lambda=0}=g^F_{\mu\nu}(X,\dot{X})\delta^{\mu}_{\rho}$ and that $(\omega_{RO})^{\mu}_{\rho}:=\frac{d(\Lambda_{RO})^{\mu}_\rho (\lambda)}{d\lambda}|_{\lambda=0}$. Then, the infinitesimal generators of the Randers-observer transformations satisfy the equation
\begin{equation}
\begin{split}
\label{randersobserverkillingequation}
g^{F}_{\mu\rho}(x,\dot{x})(\omega_{RO})^{\rho}_{\nu} + g^{F}_{\nu\rho}(x,\dot{x})(\omega_{RO})^{\rho}_{\mu}&+2C_{\mu\nu\rho}(x,\dot{x})(\omega_{RO})^{\rho}_{\sigma}\dot{x}^{\sigma}+\\
&+\frac{\partial g^{F}_{\mu\nu}(x,\dot{x})}{\partial x^{\rho}}(\omega_{RO})_{\sigma}^{\rho}x^{\sigma}=0,
\end{split}
\end{equation}
where $C_{\mu\nu\rho}(x,\dot{x}):=\frac{1}{2}\frac{\partial g^{F}_{\mu\nu}(x,\dot{x})}{\partial \dot{x}^{\rho}}$ is the so-called Cartan tensor, which measures the directional dependence of the Finsler metric \cite{chern}. Note that even for a constant background vector $a^\mu$, where the last term in Eq.\eqref{randersobserverkillingequation} vanishes, the presence of the Cartan tensor still provides noninertial directional-dependent effects. The Eq.\eqref{randersobserverkillingequation} shows how the change from one coordinate frame to another is deformed by the background vector.


\subsection{Particle transformations}
\label{section3}

In this subsection we study the symmetry transformations which preserves the particle action along the particle trajectory for a given coordinate system.  In order to do it, we start studying the symmetries arising from the Finslerian equation of motion. Then, we analyse the Killing equation for a constant background vector and find Randers transformations which are extended by local Randers transformations through the vielbein formalism.

Consider an active transformation upon a particle using the same coordinate system. The particle transformation $\Lambda_{RP}$
takes the particle from $x$ to $x'$, by
\begin{equation}
x'^{\rho}:=(\Lambda_{RP})^{\rho}_{\sigma}x^{\sigma}.
\end{equation}
Infinitesimally, we can define the displacement vector $\xi^{\rho}d\tau:=\delta x^{\rho}=(\omega_{RP})_{\sigma}^{\rho}x^{\rho}$.
We are interested in infinitesimal particle displacements $\xi^\rho$ which preserves the Randers-Finsler interval, and hence, the particle action. First, note that from the Finslerian action Eq.\eqref{finsleraction}, the equation of motion for which the Finslerian norm of $U^{F\mu}$ is constant has the form
\begin{equation}
\label{equationofmotion}
\ddot{x}^{\mu}+G^\mu (x,\dot{x})=0,
\end{equation}
where the inertial force, known as the geodesic spray coefficients,
is defined by $G^\mu = \gamma^{F\mu}_{\rho\sigma}(x,\dot{x})\dot{x}^{\rho}\dot{x}^\sigma$, and the Finslerian Christoffel symbol is given by $\gamma^{F\mu}_{\nu\rho}(x,\dot{x}):=\frac{g^{F\mu\sigma}(x,\dot{x})}{2}\Big[\partial_{\nu}g^{F}_{\sigma\rho}(x,\dot{x})+\partial_{\rho}g^{F}_{\sigma\nu}(x,\dot{x})-\partial_{\sigma}g^{F}_{\nu\rho}(x,\dot{x})\Big]$. For the Randers-Finsler spacetime, the inertial force
is given by
\begin{equation}
G^\mu=\zeta \eta^{\mu\nu}F_{\nu\sigma}U^{F\sigma} -m\zeta\left((\partial_\sigma a_{\nu})\dot{x}^\nu \dot{x}^\sigma + F_{\sigma\nu}a^{\nu} \right) U^\mu,
\end{equation}
where $F_{\mu\nu}:=\zeta(\partial_\mu a_\nu - \partial_\nu a_\mu)$ is a sort of field strength for the background vector $a_\mu$. For a constant $a^\mu$, the inertial force $G^\mu$ vanishes and then, the Finsler momentum $P^{F\mu}$ is constant. This is an expected result since the constancy of $a^\mu$ means the Randers-Finsler spacetime is homogeneous and hence, the Finsler momentum should be conserved. In this case, any translation
displacement $\xi^{\rho}$ is a generator of a symmetry of the action.

Defining a Finslerian connection we can show that the Finslerian momentum is covariantly constant along the particle worldline. Since the geometry, and hence the physics, is dependent on both the the position and velocities (and constrains among them), the best approach to Finsler geometries is work on the tangent bundle, i.e., $TM=\{(x,\dot{x})\}$ \footnote{The tangent bundle can be regarded as a configuration space and then, the Finsler geometry as a Lagrangian geometry \cite{miron}.}. Consider the orthogonal basis for $TM$,
$\delta_{\mu}:=\frac{\partial}{\partial x^{\mu}} - N_{\mu}^{\nu}\frac{\partial}{\partial \dot{x}^\nu}$, known as the horizontal derivative and the vertical derivative $\bar{\partial
}_{\mu}:=F(x,\dot{x})\frac{\partial}{\partial \dot{x}^\mu}$, where $N^{\mu}_{\nu}:=\frac{\partial G^{\mu}}{\partial \dot{x}^\nu}$ \cite{chern,miron}. The horizontal derivative measures changes when the particle moves from one point to another whereas the verticle derivative accounts for velocities changes at the same point. The horizontal covariant derivative of the Finsler metric tensor is defined as \cite{chern}
\begin{equation}
g^{F}_{\mu\nu|\rho}:= \nabla^F_\rho g^{F}_{\mu\nu}= \delta_{\rho}g^{F}_{\mu\nu}-\Gamma^{F\sigma}_{\rho\mu}g^{F}_{\sigma\nu}-\Gamma^{F\sigma}_{\rho\nu}g^{F}_{\sigma\mu}
\end{equation}
Among the many Finslerian connections which are horizontal compactible with the Finsler metric, i.e., $g^{F}_{\mu\nu|\rho}\equiv 0$ we choose the Cartan connection $\omega^{C\mu}_{\nu}:=\Gamma^{F\mu}_{\nu\sigma}dx^{\sigma} + C^{\mu}_{\nu\sigma}\delta y^{\mu}$, where $\delta y^\mu := \frac{(dy^\mu + N^{\mu}_{\nu}dx^\nu)}{F}$ and \cite{chern,rund, pfeifer}
\begin{equation}
\label{cartanconnection}
\Gamma^{F\mu}_{\nu\rho}(x,\dot{x}):=\frac{g^{F\mu\sigma}(x,\dot{x})}{2}\Big[\delta_{\nu}g^{F}_{\sigma\rho}(x,\dot{x})+\delta_{\rho}g^{F}_{\sigma\nu}(x,\dot{x})-\delta_{\sigma}g^{F}_{\nu\rho}(x,\dot{x})\Big].
\end{equation}
Using the Cartan connection, the Finslerian unit 4-velocity $U^{F\mu}:=\frac{dx^{\mu}}{d\tau_{F}}$, and hence the Finsler momentum
$P^{F\mu}:=mU^{F\mu}$ are covariantly constant \cite{chern}.
Therefore, the local isometry given by $\xi^{\rho}:=U^{F\mu}$ makes the Finslerian 4-momentum $P^{F\rho}$ the generator of the particle time evolution whose charge is the rest mass $m=g^{F}_{\mu\nu}(x,\dot{x})P^{F\mu}\xi^{\rho}$.

The covariant conservation of the momentum guarantees that the particle energy-momentum tensor $T^{F\mu\nu}:=\int_{a}^{b}{P^{F\mu}U^{F\nu}\delta^4 (x-x(\tau_F))d\tau_F}$ is horizontally conserved, $T^{F\mu\nu}_{|\nu}=0$. Defining the Rander-Finsler angular momentum tensor $J^{F\mu\nu}=x^{\mu}P^{F\nu}-x^{\nu}P^{F\mu}=\frac{\alpha(P)}{F(P)}J^{\mu\nu}$,
the spin vector $S_{\mu}^F:=\frac{1}{2}\epsilon_{\mu\nu\rho\sigma}J^{F\nu\rho}U^{F\sigma}$,
is horizontally covariantly conserved, i.e., $\frac{dS^F_\mu}{d\tau_F}=0$. Therefore, the particle has two conserved quantities, namely, the Finslerian momentum and spin vectors.

The general way to find the local symmetries $\xi^{\rho}$ of the action is through the Lie derivative of the Finsler metric, given by \cite{rund,bejancu,girelli,pfeifer}
\begin{eqnarray}
\mathcal{L}{_\xi}g^F_{\mu\nu}(x,\dot{x})&:=&\xi^{\rho}g^{F}_{\mu\nu|\rho}(x,\dot{x}) + g^{F}_{\mu\rho}(x,\dot{x})\xi^{\rho}_{|\nu}+g^{F}_{\nu\rho}(x,\dot{x})\xi^{\rho}_{|\mu}+\\
&+& 2C_{\mu\nu\rho}(x,\dot{x})\xi^{\rho}_{|\sigma}U^{F\sigma}\nonumber.
\end{eqnarray}
For a horizontal compatible connection, i.e., for $g^{F}_{\mu\nu|\rho}\equiv 0$, a Finslerian Killing vector satisfies $\mathcal{L}{_\xi}g^F_{\mu\nu}(x,\dot{x})\equiv 0$ and then
\begin{equation}
\label{finslerkillingequation}
g^{F}_{\mu\rho}(x,\dot{x})\xi^{\rho}_{|\nu}+g^{F}_{\nu\rho}(x,\dot{x})\xi^{\rho}_{|\mu}+2C_{\mu\nu\rho}(x,\dot{x})\xi^{\rho}_{|\sigma}U^{F\sigma}=0.
\end{equation}
It is worthwhile to say that using $\xi^{\rho}:=\omega^{F\rho}_{\sigma}x^{\sigma}$, the Finslerian Killing equation \eqref{finslerkillingequation} can be rewritten as Eq.\eqref{randersobserverkillingequation}. Therefore, the infinitesimal generators of the particle Randers-Finsler transformations are the same of those of the observer Randers-Finsler transformations.

Since we are interested in active particles transformations, let us consider the displacement vector in the direction of the particle motion, i.e., $\xi^{\rho}=U^{F\rho}$. The 4-acceleration is orthogonal to the 4-velocity and then, the Killing equation turns to
\begin{equation}
\label{particlekillingequation}
\xi_{\mu|\nu}+\xi_{\nu|\mu}=0.
\end{equation}
The horizontal derivatives provides a rich anisotropy dependence for the Killing fields $\xi^{\rho}$ in Eq.\eqref{particlekillingequation}.

Let us first consider the case where $a^\mu$ is constant and then the anisotropic spacetime is homogeneous.
The covariant horizontal derivative turns into $\xi_{\mu|\nu}=\delta_{\nu}\xi_{\mu}$ and a constant 4-vector $\epsilon^\mu$ satisfies the equation \eqref{particlekillingequation}. Thus, we can use the Killing vector $\epsilon^\mu$ to define a translation operation by
\begin{equation}
T(\lambda):=e^{i(\epsilon^{\mu}P^{F}_{\mu})\lambda},
\end{equation}
where the translation generator is defined as
\begin{equation}
P_{\mu}^{F}:=-i\delta_{\mu}.
\end{equation}
The infinitesimal translation is $\xi^\rho=\epsilon^{\mu}\delta_{\mu}x^{\rho}=\epsilon^\rho$.

In addition, note that defining $\omega_{\mu\nu}:=\delta_{\nu}\xi_{\mu}$, Eq.\eqref{particlekillingequation} can be written as $\omega_{\mu\nu}+\omega_{\nu\mu}=0$, where
\begin{equation}
\omega^{F\mu}_{\nu}:=g^{F\mu\rho}(\dot{x})\omega_{\rho\nu}.
\end{equation}
The Killing vector related to this symmetry can be defined as
\begin{equation}
\xi^\rho=\omega^{F\rho}_{\sigma}(\dot{x})x^{\sigma}.
\end{equation}
Defining the proper Rander-Finlser generators
\begin{equation}
J^{F\mu\nu}:=g^{F\mu\rho}(x,\dot{x})g^{F\nu\sigma}(x,\dot{x})J^F_{\rho\sigma},
\end{equation}
where
\begin{equation}
J^{F}_{\mu\nu}	:=	x_{\mu}P_{F\nu}-x_{\nu}P_{F\mu} = -i(x_{\mu}\delta_{\nu}-x_{\nu}\delta_{\mu}),
\end{equation}
the proper Randers-Finsler particle transformation can be defined as
\begin{equation}
\Lambda_{RP}(\lambda):=e^{i\frac{1}{2}(M^{F\mu\nu}\omega_{\mu\nu})\lambda}.
\end{equation}
Then, the infinitesimal proper Randers-Finsler particle transformation takes the form
\begin{eqnarray}
\xi^\rho	&	=	&	-(\omega^{F\mu\nu}x_{\mu}\delta_\nu)x^{\rho}\\
			&	=	&	\frac{\alpha}{F}\Big\{\omega^\rho_\nu - \zeta\frac{\alpha}{F}\Big[U^\rho a_\nu  +U^\mu a^\rho \omega_{\mu\nu}\Big]\nonumber\\
			&	+	&	 \left(\frac{\alpha}{F}\right)^2 [\zeta (a\cdot U)+\zeta^2 a^2]U^\rho U_\mu \omega^{\mu}_{\nu}\Big\}x^\nu\nonumber,
\end{eqnarray}
which is a combination of the the Lorentz infinitesimal transformation $\omega^\rho _\nu$ and its action on the Lorentzian 4-velocity $U^\rho$. It is worthwhile to say that, since we are employing a particle transformation, the background vector $a^\rho$ is left unchanged under this action. Expanding the infinitesimal proper Randers-Finsler particle transformation in powers of $\zeta$, we obtain $\delta x^\rho = \xi^\rho = \Big\{\omega^{\rho}_\sigma - \zeta \Big[(a\cdot U)\omega^{\rho}_\sigma + U^\rho a_{\sigma} + a^\rho U^\mu \omega_{\mu\sigma} - (a\cdot U) U^\rho \omega_{\sigma}^{\mu}U_\mu \Big] + \mathcal{O}(\zeta^2)\Big\}x^{\sigma}$.

Likewise the linear momentum operator \eqref{randersmomentumcovariant}, the Randers-Finsler generators are related to the Lorentz generators by
\begin{eqnarray}
J^F_{\mu\nu} 	&	=	&	J_{\mu\nu}+\zeta C_{\mu\nu}(x,a,m),
\end{eqnarray}
where $C_{\mu\nu}(x,a):=-m(x_{\mu}a_\nu - x_\nu a_\mu)$.
The anisotropic Randers-Lorentz parameters $\omega^{F\mu\nu}:=g^{F\mu\rho}g^{F\nu\sigma}\omega_{\rho\sigma}$ can be rewritten as $\omega^{F\mu\nu}=\omega^{\mu\nu}+\zeta \tilde{\omega}^{\mu\nu}(\zeta,a,U)$, where $\tilde{\omega}^{\mu\nu}$ contains all the anisotropic terms. Thus,
\begin{equation}
\omega^{F\mu\nu}J^F_{\mu\nu}=\omega^{\mu\nu}J_{\mu\nu}+\zeta(\omega^{\mu\nu}C_{\mu\nu} + \tilde{\omega}^{\mu\nu}J_{\mu\nu}) + \zeta^2 \tilde{\omega^{\mu\nu}}C_{\mu\nu},
\end{equation}
and hence, the particle Randers-Finsler transformation can be written as
\begin{equation}
\Lambda_{RP}(\lambda)=\Lambda\circ \Sigma \circ \Theta,
\end{equation}
where $\Lambda=e^{\frac{i}{2}\omega^{\mu\nu}J^F{\mu\nu}}$, $\Sigma:=e^{\frac{i}{2}\zeta(\omega^{\mu\nu}C_{\mu\nu} + \tilde{\omega}^{\mu\nu}J^F{\mu\nu})}$ and $\Theta:=e^{\frac{i}{2}(\zeta^2 \tilde{\omega^{\mu\nu}}C_{\mu\nu})}$. The operators $\Sigma$ and $\Theta$ are responsible for the mass shell deformation.

For a varying background vector, the Randers-Finsler metric also depends on the position. By applying the vielbein formalism, we can rewrite the Randers metric \eqref{randersmetric} as \cite{chern}
\begin{equation}\label{metricfactorization}
g_{\mu\nu}^F=\Lambda^{Fb}_\mu (x,\dot{x})\Lambda^{Fc}_\nu (x,\dot{x})\eta_{bc},
\end{equation}
where the vielbeins are given by
\begin{eqnarray}
\label{randersvielbein}
\Lambda^{Fb}_\mu (x,\dot{x})&=&\sqrt{\frac{F}{\alpha}}\Big\{\Lambda^{a}_\mu + \left(\frac{\alpha}{F}\right)^2 \Big[-\frac{\beta}{2\alpha}U^\mu U_\mu +U^\mu a_b + a^\mu U_b + a^\mu a_b \Big] \Big\}\nonumber
\end{eqnarray}
Note the presence of the anisotropic factor $\sqrt{\frac{F}{\alpha}}$, a feature also exhibited by the Bogoslovsky spacetime \cite{bogoslovsky}. The vielbein $\Lambda^{Fb}_\mu (x,\dot{x})$
can be understood as a local Randers transformation. Therefore, both the infinitesimal $\xi^{\rho}$ and local Randers transormation $\Lambda^{Fb}_\mu (x,\dot{x})$ can be regarded as deformed Lorentz transformations.


\section{Randers algebra}
\label{section4}

The Finsler algebra is strongly modified by the anisotropy of the
action. In fact, the momentum algebra is given by
\begin{eqnarray}
[P^F_\mu , P^F_\nu]	&	=	&	-[\delta_\mu , \delta_\nu]=2 \delta_{[\mu} N^\rho_{\nu]}\frac{\partial}{\partial \dot{x}^\rho}\nonumber\\
					&	=	&	 R^{F\sigma}_{\rho\mu\nu}(x,P^F)P^{F\rho}\frac{\partial}{\partial P^{F\sigma}}.
\label{finslerlinearmomentumcommutator}
\end{eqnarray}
The nonvanishing commutator in Eq.\eqref{finslerlinearmomentumcommutator} can be understood as a result of the anisotropy yields to non-inertial effects measures with the Finslerian horizontal curvature $R^{F\sigma}_{\rho\mu\nu}(x,\dot{x})$. A special case is given by the horizontal flat Finslerian spacetimes, where $R^{F\sigma}_{\rho\mu\nu}(x,\dot{x})=0$.
In Randers-Finsler spacetimes, the linear momentum algebra is modified at the leading order by
\begin{equation}
[P^F_\mu , P^F_\nu]	= \eta^{\rho\sigma}\Big\{\delta_\mu \left(\frac{\alpha}{F}F_{\sigma\nu}\right)-\delta_\nu \left(\frac{\alpha}{F}F_{\sigma\mu}\right)\Big\}\frac{\partial}{\partial P^{F\rho}}.	
\end{equation}
Therefore, as long as the vector field $a_\mu$ satisfies the Randers-Finsler flat condition
\begin{equation}
\label{randersflatcondition}
2\delta_{[\mu}\left(\frac{\alpha}{F}F_{\nu]\sigma}\right)=0,
\end{equation}
the linear momentum generator satisfies $[P^F_\mu , P^F_\nu]=0$ at leading-order. The constant $a_\mu$ configuration clearly satisfies Eq.\eqref{randersflatcondition}.

The angular momentum generators are more sensible to the anisotropic structure. In fact, for a constant $a^\mu$, then the  angular momentum generators satisfy
\begin{eqnarray}
\label{finslerangularmomentumanglebra}
[J^F_{\mu\nu} , J^F_{\lambda\rho}]	&	=	&	-g^F_{\mu\lambda}(P^{F})J^F_{\nu\rho}-g^F_{\nu\rho}(P^{F})J^F_{\mu\lambda}\nonumber\\
									&	+	&g^F_{\mu\rho}(P^{F})J^F_{\nu\lambda} + g^F_{\nu\lambda}(P^{F})J^F_{\mu\rho}.
\end{eqnarray}
We can rewrite Eq.\eqref{finslerangularmomentumanglebra} as $[J^F_{\mu\nu} , J^F_{\lambda\rho}]=
\frac{\alpha}{F}(-\eta_{\mu\lambda}J^F_{\nu\rho}-\eta_{\nu\rho}J^F_{\mu\lambda} +	\eta_{\mu\rho}J^F_{\nu\lambda} + \eta_{\nu\lambda}J^F_{\mu\rho}) -D^F_{\mu\lambda}(P^{F})J^F_{\nu\rho}-D^F_{\nu\rho}(P^{F})J^F_{\mu\lambda} D^F_{\mu\rho}(P^{F})J^F_{\nu\lambda} + D^F_{\nu\lambda}(P^{F})J^F_{\mu\rho}$,
where $D_{\mu\nu}^F:=g_{\mu\nu}^F-\frac{\alpha}{F}\eta_{\mu\nu}$.
Assuming the background vector is still constant, the commutator between $P^F_\mu$ and $J^F_{\nu\lambda}$ takes the form
\begin{eqnarray}
\label{thirdcommutator}
[P^F_\mu, J^F_{\nu\lambda}]	&	=	&	g^F_{\mu\nu}(P^{F})P^F_\lambda - g^{F}_{\mu\lambda}(P^{F})P^F_\nu.
\end{eqnarray}

It is worthwhile to say that the commutation relations in equations \eqref{finslerlinearmomentumcommutator},\eqref{finslerangularmomentumanglebra} and \eqref{thirdcommutator} have an analogous form of that of the Lorentz invariant transformations, changing only $J^F_{\mu\nu}\rightarrow J^F_{\mu\nu}$ and $\eta_{\mu\nu}\rightarrow g^F_{\mu\nu}(P^{F})$.

The Finsler metric deforms the algebra between momentum, Randers transformations and position. Indeed, for a constant $a^\mu$,
\begin{equation}
[P^F_\mu , x_\nu]=-i\left(g^F_{\mu\nu}(x,P^F)+\delta_\mu(g^F_{\nu\rho}(x,P^F)x^\rho\right),
\end{equation}
\begin{equation}
[J^{F}_{\mu\nu} , x_{\lambda} ]=-i\Big\{x_\mu g^F_{\nu\lambda}(x,P^F)-x_\nu g^F_{\mu\lambda}(x,P^F)\Big\}
\end{equation}
Since for $\zeta=0$, $g^F_{\mu\nu}(x,P^F)=\eta_{\mu\nu}$, the Randers algebra resembles the deformed Heisenberg-Weyl algebra of the $\kappa$-Minkowski spacetime \cite{kalgebra}.

From the Randers-Finsler generators, we can define the Randers-Finsler boost generators
\begin{eqnarray}
K_i^F	&	:=	&	J^F_{0i}=K_i + \zeta m (x_i a_0 - x_0 a_i),
\end{eqnarray}
and the Randers-Finsler angular momentum generator
\begin{eqnarray}
J^F_i	&	:=	&	-\frac{1}{2}\epsilon^{jk}_{i}J^F_{jk} =J_i - \zeta m \epsilon^{jk}_i a_j x_k.
\end{eqnarray}
Remarkably, the background vector $a^\mu$ produces not only an expected background boost generator but a angular momentum generator as well.

At the particle rest frame where $g^F_{00}(\dot{x})=1$, the boosts and angular momentum generators satisfy
\begin{equation}
[K^F_i , K^F_j]=\epsilon^k_{ij}J^F_k + 2 g^{F}_{0[i}K^F_{j]}
\end{equation}
\begin{equation}
[K^F_i , J^F_j]=\epsilon^k_{ij}K^F_k + 2 g^{F}_{0[i}J^F_{j]}
\end{equation}
\begin{equation}
[J^F_i , J^F_j]=\epsilon^k_{ij}J^F_k
\end{equation}
For a timelike background vector $a^\mu$, where we can set $g^F_{0i}(\dot{x})=0$, the Randers algebra takes a form of a
extension of the Lorentz algebra by $K_\mu \rightarrow K^F_\mu$ and $J_{\mu\nu} \rightarrow J^F_{\mu\nu}$.
For a spacelike and lighlike $a^\mu$, the presence of the terms $g^F_{0i}$ provides mixtures between boosts and Randers generators.

Besides the boosts and angular momentum operators, we can define a Casimir-Finsler operator as
\begin{equation}
\label{casimir}
P^{F2}:=g^{F\mu\nu}(P^F)P^F_\mu P^F_\nu.
\end{equation}
The Casimir operator in Eq.\eqref{casimir} is an scalar operator generically nonquadratic.
For a constant $a^\mu$, $[P^{F2},P^F_\mu]=[P^{F2},J^F_{\mu\nu}]=0$,
and hence, the Finsler Casimir operator is preserved under translations and Randers-Finsler transformations.
For the Randers-Finsler spacetime, the Casimir operator takes the form
\begin{equation}
P^{F2}=\eta^{\mu\nu}P^F_\mu P^F_\nu -2\zeta a^\rho P^F_\rho \sqrt{-\eta^{\mu\nu}P^F_\mu P^F_\nu} - \zeta^2 a^\mu a^\nu
P^F_\mu P^F_\nu.
\end{equation}

In addition, we can define a Pauli-Lubanski-Finsler vector operator by
\begin{eqnarray}
\label{paulilubanski}
W^{F\mu}	&	:=	&	\frac{1}{2}\epsilon^{\mu\nu\lambda\rho}P^F_\nu J^F_{\lambda\rho}\nonumber\\
			&	=	&	W^\mu +\frac{\zeta}{2}\epsilon^{\mu\nu\lambda\rho}[P_\nu C_{\lambda\rho}-m a_\nu J^F_{\lambda\rho}]-\frac{\zeta^2 m}{2}\epsilon^{\mu\nu\lambda\rho}a_\nu C_{\lambda\rho}.
\end{eqnarray}
The Pauli-Lubanski-Finsler operator satisfies the orthogonality condition $P^F_{\mu}W^{F\mu}=0$ and the commutation relations
$[W^F_{\mu} , P^F_{\nu}]=0$, $[W^F_\mu, J^F_{\nu\lambda}]=g^F_{\mu\nu}(P^{F})W^F_\lambda - g^{F}_{\mu\lambda}(P^{F})W^F_\nu$.


\section{A Randers field theory}
\label{section5}

Once analysed the symmetries of the Randers-Finsler spacetimes, let us define a dynamics for Randers symmetric fields.
Consider a Randers invariant action
\begin{equation}
S=\int_{M}{d^{4}x \mathcal{L}}.
\end{equation}
Likewise the particle action, we are interested in actions build from Randers scalars, i.e., invariant over
Randers transformations. The relation $P_{\mu}^F=-i\delta_{\mu}$ suggests an approach where the direction-dependence of the geometry becomes a momentum-dependence of the metric. Assuming the fields have only position dependence, the momentum operator
has its origin on position variations, i.e., $P^F_{\mu}=-i \nabla^F_{\mu}$. Thus, the Finsler metric can be regarded as a differential operator, where $g^F_{\mu\nu}(x,y)\rightarrow g^F_{\mu\nu}(x,\nabla^F)$ and $y^\mu \rightarrow \nabla^{F\mu}$.
The relation $P^{F\mu}=\frac{\alpha}{F}P^\mu$ and the homogeneity of the Finsler metric, allow us to write the Finsler metric
as the operator $g^F_{\mu\nu}(x,\nabla)$. This approach of considering the Finsler metric as an operator is similar to the noncanonical kinetic terms \cite{noncanonicalkinetic}. Unlike the field theories defines on the tangent bunblde $TM_4$ \cite{pfeifer}, the Finsler metric operator procedure enable us to propose a field theory defined on the spacetime $M_4$ itself.

\subsection{Scalar field}

A scalar field $\Phi(x)$ is naturally invariant over the Randers transformations, i.e., $\Phi(\Lambda_R x)=\Phi(x)$.
We propose a Randers invariant action for the scalar field is given by
\begin{equation}
\label{finslerscalaraction}
S_{\Phi}:=-\frac{1}{2}\int_{M}{d^{4}x\{\nabla_{\mu}\Phi K^{\mu\nu}(x,\nabla) \nabla_{\nu}\Phi + m^2 \sqrt{-g^F(x,\nabla)} \Phi^2]\}}.
\end{equation}
where
\begin{equation}
K^{\mu\nu}:=\sqrt{-g^F(x,\nabla)}g^{F\mu\nu}(x,\nabla).
\end{equation}
For $\zeta=0$, i.e., for a Local Lorentz symmetric spacetime, the action \eqref{finslerscalaraction} yields to a minimal coupling of the scalar field in a curved spacetime.

In the Randers spacetime, by means of the identification $y^\mu\rightarrow \nabla^\mu$, the contravariant metric has the form \cite{chern}
\begin{eqnarray}
g^{F\mu\nu}(x,\nabla)&=&\frac{1}{1+\zeta^2 a\cdot\nabla} g^{\mu\nu}-\zeta^2\frac{1}{(1+\zeta^2 a\cdot\nabla)^2} (a^\mu \nabla^\nu + a^\nu \nabla^\mu)\\
&+&\zeta^4\frac{1}{(1+\zeta^2 a\cdot\nabla)^3} \Big[a\cdot\nabla+ a^2\Big]\nabla^\mu \nabla^\nu \nonumber,
\end{eqnarray}
where $a\cdot\nabla:=a^\mu \nabla_\mu$. Then, defining a dimensionless background field $b^\mu:=\zeta a^\mu$, the Randers action for the scalar field can be rewritten as
\begin{equation}
\label{randersscalaraction1}
\begin{split}
S_{\Phi}=-\frac{1}{2}\int_{M}d^{4}x \sqrt{-g}\Bigg\{& g^{\mu\nu}\nabla_\mu \Phi [(1+\zeta b\cdot\nabla)^{\frac{3}{2}}]\nabla_\nu \Phi + (1+\zeta^2 a\cdot\nabla)^{\frac{5}{2}}m^2 \Phi^2 + \\
&\zeta\nabla_\mu \Phi \Big[(1+\zeta b\cdot\nabla)^{\frac{3}{2}}(g^{\mu\rho}b^\nu + g^{\nu\rho}b^\mu)\Big]\nabla_\rho \nabla_\nu \Phi + \\
& \zeta^2 g^{\mu\rho}g^{\nu\sigma}\nabla_\mu \Phi \Big[\frac{(\zeta \beta + b^2)}{(1+\zeta b\cdot\nabla)^{\frac{1}{2}}}\Big]\nabla_\nu \nabla_\rho \nabla_\sigma \Phi \Bigg\} \nonumber.
\end{split}
\end{equation}
The action exhibits nonlocal dynamical terms which for the Minkwoski background spacetime, $(g_{\mu\nu}=\eta_{\mu\nu})$, are similar to those of VSR \cite{vsr}.

The perturbative character of the Randers spacetime allow us to rewrite the Randers Lagrangian as
\begin{equation}
\mathcal{L}^{F}_\phi=\mathcal{L}_{LI}+\zeta \mathcal{L}^{1}_{LV}+\zeta^2 \mathcal{L}^{2}_{LV} +  \cdots,
\end{equation}
where $\mathcal{L}_{LI}:=-\frac{\sqrt{-g}}{2}(g^{\mu\nu}\nabla_\mu \Phi\partial_\nu \Phi + m^2 \Phi)$ is the usual Lorentz-invariant Klein-Gordon Lagrangian. The first-order Lorentz-violating Lagrangian is given by
\begin{equation}
\label{scalarlvlagrangian}
\mathcal{L}^{1}_{LV}= \nabla_\mu \Phi (K^{5})^{\mu\nu}\nabla_\nu \Phi - \frac{5m^2 \sqrt{-g}}{4} b^\rho \nabla_\rho (\Phi^2),
\end{equation}
where the mass dimension five Lorentz violating operator $(K^{5})^{\mu\nu}$ has the form
\begin{equation}
(K^{5})^{\mu\nu}:=-\frac{\sqrt{-g}}{4}\Big\{3 g^{\mu\nu}b^\rho - 2 (g^{\mu\rho}b^\nu + g^{\nu\rho}b^\mu) \Big\}\nabla_\rho.
\end{equation}
The second-order terms forms the Lorentz-violating Lagrangian
\begin{eqnarray}
\label{scalarlvlagrangian2}
\mathcal{L}^{2}_{LV}= \nabla_\mu \Phi (K^{6})^{\mu\nu}\nabla_\nu \Phi - \frac{15m^2 \sqrt{-g}}{8} b^\rho b^\sigma \nabla_\rho\nabla_\sigma (\Phi^2)\nonumber,
\end{eqnarray}
where the mass dimension six operator $(K^{6})^{\mu\nu}$ is defined as
\begin{equation}
(K^{6})^{\mu\nu}:=-\frac{\sqrt{-g}}{4}\Big\{\frac{3}{8} g^{\mu\nu}b^\rho b^\sigma - \frac{b^2}{2} g^{\mu\rho}g^{\nu\sigma}\Big\}\nabla_\rho \nabla_\sigma.
\end{equation}
It is worthwhile to say that the last term in the first-order perturbed Lagrangian Eq.\eqref{scalarlvlagrangian}, for a covariantly constant background vector $b^\mu$, provides a total derivative term which can be dropped from the action.
The operators $(K^{5})^{\mu\nu}$ and $(K^{6})^{\mu\nu}$ in a flat background metric have the same form of the nonminimal Standard Model Extension for dimension 5 and 6 Lorentz violating operators.


\subsection{Vector Gauge field}

Now let us analyse the dynamics of a vector field invariant over the Randers transformations $A'^{\mu}=\Lambda^{\mu}_{R\nu}(x)A^{\nu}$. We propose a Finslerian extension of the field strength using the horizontal covariant derivatives instead of the Levi-Civita covariant derivative, i.e.,
\begin{equation}
F^{R}_{\mu\nu}:=\nabla^F_\mu A_\nu - \nabla^F_\nu A_\mu = A_{\nu|\mu}-A_{\mu|\nu},
\end{equation}
where
$\nabla^F_\mu A_\nu:=\delta_\mu A_\mu - \Gamma^{F\rho}_{\mu\nu}A_\rho$ and $F'_{\mu\nu}=\Lambda^{t\rho}_{R\mu}\Lambda^{t\sigma}_{R\nu}F_{\rho\sigma}$.

Using the Cartan connection, the Finslerian field strength can be rewritten as $F^{R}_{\mu\nu}=\delta_{[\mu}A_{\nu]}$.
Under a gauge transformation $A'_\mu=A_\mu + \delta_\mu f$, the field strength changes as $F'^{R}_{\mu\nu}=F^R_{\mu\nu}+[\delta_\mu , \delta_\nu]f$. Since in our approach, $A=A(x)$, then $\delta_\mu = \partial_\mu$ and $F^{R}_{\mu\nu}=F_{\mu\nu}$. Therefore, the electric and magnetic components of the field strength and the gauge symmetry are preserved in Randers spacetime.

A gauge Finslerian extension of the Maxwell action has the form
\begin{eqnarray}
\label{randersgaugeaction}
S^F_{A}&:=&-\frac{1}{4}\int d^4 x\Big\{F_{\mu\nu}K^{F\mu\nu\rho\sigma}(x,\nabla)F_{\rho\sigma}\Big\},
\end{eqnarray}
where
\begin{equation}
K^{F\mu\nu\rho\sigma}:=\sqrt{-g^F(x,\nabla)}g^{F\mu\rho}(x,\nabla)g^{F\nu\sigma}(x,\nabla).
\end{equation}
Note that for $\zeta=0$, the tensor $K^{F\mu\nu\rho\sigma}$ turns into $\sqrt{-g(x)}g^{\mu\rho}(x)g^{\nu\sigma}(x)$ we obtain the usual Maxwell term $\mathcal{L}_A=-\frac{\sqrt{-g(x)}}{4}g^{\mu\rho}(x)g^{\nu\sigma}(x)F_{\mu\nu}F_{\rho\sigma}$.

In Randers spacetime, the Finsler gauge action \eqref{randersgaugeaction} yields to the Finsler gauge Lagrangian
\begin{eqnarray}
\label{randersgaugelagrangian}
\mathcal{L}^{F}_{A}&=&-\frac{\sqrt{-g}}{4}F_{\mu\nu}\Big\{(1+\zeta b\cdot\nabla)^{\frac{1}{2}}g^{\mu\rho}g^{\nu\sigma} -2\zeta^2 \Big[\frac{g^{\mu\rho}g^{\nu\lambda}g^{\sigma\xi}}{(1+\zeta b\cdot \nabla)^{\frac{1}{2}}}(\nabla_\lambda b_\xi + \nabla_\lambda b_\xi)\Big]\\
&+& 2\zeta^2 \Big[\frac{g^{\mu\rho}g^{\nu\lambda}g^{\sigma\xi}(\zeta \beta + b^2)}{(1+\zeta b\cdot \nabla)^{\frac{3}{2}}}\nabla_\lambda \nabla_\xi \Big]\Big\}F_{\rho\sigma}\nonumber.
\end{eqnarray}
The Finsler gauge action in Eq.\eqref{randersgaugelagrangian} exhibits nonlocal operators. Let us analyse the gauge Lagrangian
in Eq.\eqref{randersgaugelagrangian} term by term.
The first term
\begin{equation}
\mathcal{L}^1_{A}=-\frac{\sqrt{-g}}{4}F_{\mu\nu}(1+\zeta b\cdot\nabla)^{\frac{1}{2}}g^{\mu\rho}g^{\nu\sigma}F_{\rho\sigma},
\end{equation}
can be expanded in powers of $\zeta$ as
\begin{equation}
\mathcal{L}^1_{A}=-\frac{\sqrt{-g}}{4}g^{\mu\rho}g^{\nu\sigma}F_{\mu\nu} F_{\rho\sigma} + \zeta F_{\mu\nu}(\hat{k}^{(5)}_F)^{\mu\nu\rho\sigma}F_{\rho\sigma}+\zeta^2 F_{\mu\nu}(\hat{k}^{(6)}_F)^{\mu\nu\rho\sigma}F_{\rho\sigma} + \cdots ,
\end{equation}
where the zero order term consists of the usual Maxwell Lagrangian and the first-order and second-order Lorentz-violating Lagrangian $\mathcal{L}_{ALV}$ term are respectively, $(\hat{k}^{(5)}_F)^{\mu\nu\rho\sigma}:=-\frac{\sqrt{-g}}{8}g^{\mu\rho}g^{\nu\sigma}b^\lambda \nabla_\lambda$ and $(\hat{k}^{(6)}_F)^{\mu\nu\rho\sigma}:=-\frac{\sqrt{-g}}{8}g^{\mu\rho}g^{\nu\sigma}b^\lambda b^\xi \nabla_\lambda \nabla_\xi$. For a flat background metric, $g_{\mu\nu}=\eta_{\mu\nu}$, these terms can be regarded as the dimension five and six Lorentz violating operators belonging to the nonminimal SME \cite{gaugesmenonminimal}.

The second term
\begin{equation}
\mathcal{L}^2_{A}:=\frac{\sqrt{-g}}{2}\zeta^2 F_{\mu\nu}\Bigg[\frac{g^{\mu\rho}g^{\nu\lambda}g^{\sigma\xi}}{(1+\zeta b\cdot \nabla)^{\frac{1}{2}}}(\nabla_\lambda b_\xi + \nabla_\lambda b_\xi)\Bigg]F_{\rho\sigma},
\end{equation}
can be expanded as
\begin{equation}
\mathcal{L}^2_{A}=\zeta^2 [F_{\mu\nu}(\hat{k}_F^{5})^{\mu\nu\rho\sigma}F_{\rho\sigma}+F_{\mu\nu}\tilde{k}^{\mu\nu\rho\sigma}F_{\rho\sigma}]+\cdots,
\end{equation}
where $(\hat{k}_F^{5})^{\mu\nu\rho\sigma}:=\frac{\sqrt{-g}}{2}g^{\mu\rho}(g^{\nu\lambda}b^\sigma  + g^{\sigma\lambda}b^\nu)\nabla_\lambda$ and $\tilde{k}^{\mu\nu\rho\sigma}:=\frac{\sqrt{-g}}{2}g^{\mu\rho}[(\nabla^\nu b^\sigma + \nabla^\sigma b^\rho)]$. It is worth to note that $\tilde{k}^{\mu\nu\rho\sigma}=0$ for a covariantly constant background vector, i.e., for a Randers spacetime of Berwald type. For a background flat spacetime and constant background vector, the mass dimension five operator
$(\hat{k}_F^{5})^{\mu\nu\rho\sigma}$ belongs to the nonminimal SME \cite{gaugesmenonminimal}.

The third term
\begin{equation}
\mathcal{L}^2_{A}:=\frac{\sqrt{-g}}{2}\zeta^2 F_{\mu\nu}\Bigg[\frac{g^{\mu\rho}g^{\nu\lambda}g^{\sigma\xi}(\zeta \beta + b^2)}{(1+\zeta b\cdot \nabla)^{\frac{3}{2}}}\nabla_\lambda \nabla_\xi \Bigg]F_{\rho\sigma},
\end{equation}
can be rewritten as
\begin{equation}
\mathcal{L}^3_{A}:=\zeta^2 F_{\mu\nu}(\hat{k}_F^{(6)})^{\mu\nu\rho\sigma}F_{\rho\sigma},
\end{equation}
where $(\hat{k}_F^{(6)})^{\mu\nu\rho\sigma}:=-\frac{\sqrt{-g}}{2}b^2 g^{\mu\rho}g^{\nu\lambda}g^{\sigma\xi}\nabla_\lambda \nabla_\xi$,
for a flat background metric and vector, is a dimension six Lorentz coefficient of the nonminimal SME \cite{gaugesmenonminimal}.

Another gauge and Finsler invariant Lagrangian is given by
\begin{equation}
\label{finslergaugeaction2}
\tilde{S}^F:=\int d^4 x[\sqrt{-g^F(x,\nabla)}\epsilon^{\mu\nu\rho\sigma}F_{\mu\nu}F_{\rho\sigma}],
\end{equation}
that in Randers spacetime yields to $\mathcal{\tilde{L}}^F=(1+\zeta^2 a\cdot\nabla)^{\frac{3}{2}}\epsilon^{\mu\nu\rho\sigma}F_{\mu\nu}F_{\rho\sigma}$. Expanding this Finsler gauge Lagrangian, we obtain
\begin{equation}
\mathcal{\tilde{L}}^F=\mathcal{\tilde{L}}_{LI}+\Big[\zeta (\hat{k}^{(5)}_F)^{\mu\nu\rho\sigma}+\zeta^2 (\hat{k}^{(6)}_F)^{\mu\nu\rho\sigma} + \cdots \Big]F_{\mu\nu}F_{\rho\sigma},
\end{equation}
where $\mathcal{\tilde{L}}_{LI}=\epsilon^{\mu\nu\rho\sigma}F_{\mu\nu}F_{\rho\sigma}$ , $(\hat{k}^{(5)}_F)^{\mu\nu\rho\sigma}:=\frac{3}{2}\epsilon^{\mu\nu\rho\sigma}b^\lambda \nabla_\lambda$ and $(\hat{k}^{(6)}_F)^{\mu\nu\rho\sigma}:=-\frac{3}{4}\epsilon^{\mu\nu\rho\sigma}b^\lambda b^\epsilon \nabla_\lambda \nabla_\epsilon$.

The Randers symmetry also allows the following gauge invariant action coupling the Field strength and the gauge field
\begin{equation}
\label{finslergaugeaction3}
\hat{S}^F:=\zeta \int d^4 x[\epsilon^{\mu\nu\rho\sigma}A_\mu \hat{K}_\nu F_{\rho\sigma}],
\end{equation}
where $\hat{K}_\nu := \sqrt{-g^F(x,\nabla)} a_\nu$. Expanding the action in Eq.\eqref{finslergaugeaction3}, we obtain the Lorentz violating Lagrangian
\begin{equation}
\hat{\mathcal{L}}^F=\zeta \epsilon^{\mu\nu\rho\sigma}A_\mu a_\nu F_{\rho\sigma} + \zeta^2 \epsilon^{\mu\nu\rho\sigma}a^\lambda A_\mu \nabla_\lambda a_\nu F_{\rho\sigma}+\zeta^2 \epsilon^{\mu\nu\rho\sigma}a^\lambda A_\mu  a_\nu \nabla_\lambda F_{\rho\sigma} + \cdots,
\end{equation}
where the first-order term is the Carrol-Field-Jackiw Lagrangian \cite{Carroll}, the second term vanishes for a constant background vector in flat spacetime and the last term provides a dimension four operator.

Therefore, the Randers actions Eq.\eqref{randersgaugeaction} and Eq.\eqref{finslergaugeaction2} for the vector gauge field can be regarded as a series of Lorentz violating terms of the nonminimal Standard Model Extension \cite{gaugesmenonminimal}.


\subsection{Fermionic field}

In order to describe fermions, we can adopt the vielbein formalism. The Randers invariant gamma matrices are defined as
\begin{equation}
\gamma^{F\mu}(x,P):=E^{F\mu}_{b}(x,P)\gamma^b,
\end{equation}
where the vielbeins $E^{F\mu}_{b}(x,P)$ satisfy $g^{F\mu\nu}(x,P)=E^{F\mu}_b (x,P) E^{F\nu}_c (x,P)\eta^{bc}$. Thus $E^{F\mu}_b$ are the inverse of the vielbeins $E^{Fb}_\mu$ in Eq.\eqref{randersvielbein}. In Randers spacetime the vielbein are given by
\begin{eqnarray}
\label{randersvielbeins}
E^{F\mu}_{b}(x,P)&=&\sqrt{\frac{\alpha}{F}}\Big\{ E^{\mu}_{b}(x) + \frac{\zeta^2}{2}\frac{\alpha}{F}\Big[\frac{\alpha}{F}\left(\frac{\beta}{\alpha} + \zeta^2 a^2 \right) P^\mu P_b -(P^\mu a_b + a^\mu P_b)\Big]\Big\}.
\end{eqnarray}
Therefore, the Finslerian gamma matrices in Randers spacetime have the form
\begin{eqnarray}
\gamma^{F\mu}(x,P)&=&\sqrt{\frac{\alpha}{F}}\Bigg\{\gamma^{\mu}(x)+ + \frac{\zeta^2}{2}\frac{\alpha}{F}\Big[\frac{\alpha}{F}\left(\frac{\beta}{\alpha} + \zeta^2 a^2 \right) P^\mu \slashed{P} -(P^\mu \slashed{a} + a^\mu \slashed{P})\Big]\Big\},
\end{eqnarray}
where $\gamma^{\mu}(x):=E^\mu_b(x)\gamma^b$, $\slashed{P}:=\gamma^b P_b$ and $\slashed{a}:=\gamma^b a_b$.

Let us define the Finsler action for the fermionic field at the background flat spacetime $(g_{\mu\nu}=\eta_{\mu\nu})$.
As done for the Finsler metric, the Finsler gamma matrices are seen as operators, defined as
\begin{equation}
\hat{\gamma}^{F\mu}(D):=\sqrt{-g^{F}(D)}\gamma^{F\mu}(D).
\end{equation}
A Finsler-invariant Dirac action has the form
\begin{equation}
\label{finslerdiracaction}
S^F_{\Psi}:=\int_{M_4} d^4x\Big[\bar{\Psi}[i\hat{\gamma}^{F\mu}(D)D_\mu -m]\Psi + H.C. \Big],
\end{equation}
where the Finsler covariant derivative $D_\mu$ is defined as
\begin{equation}
D^F_\mu:=\partial_\mu +\frac{i}{2}\Gamma^{Fbc}_{\mu}\Sigma_{bc},
\end{equation}
and the local-flat generators $\Sigma_{bc}$ are defined as $\Sigma_{bc}:=\frac{i}{4}[\gamma_b , \gamma_c]$.
For a constant background metric $\eta_{\mu\nu}$, the Finslerian connection coefficients depends only on the derivatives of the background vector $b_\mu=\zeta a_\mu$. Therefore, assuming a constant background vector, we adopt $D_\mu = \partial_\mu$.
In Randers spacetime the Finsler Lagrangian for the fermion takes the form
\begin{eqnarray}
\label{randersfermionlagrangian}
\mathcal{L}^F_{\Psi}&=&\bar{\Psi}(1+\zeta b\cdot \partial)^{\frac{5}{2}}\Bigg\{\frac{i}{(1+\zeta b\cdot \partial)^{\frac{1}{2}}}\Big[\gamma^\mu \partial_\mu - \frac{\zeta}{2}\frac{(\gamma^\mu b^\nu \partial_\mu \partial_\nu + \gamma^\mu \eta^{\nu\rho}b_\mu \partial_\nu \partial_\rho)}{(1+\zeta b\cdot \partial)}+\\
 &+&\frac{\zeta^2}{2}\frac{(\zeta b\cdot \partial +b^2)}{(1+\zeta b\cdot \partial)^2}\gamma^\mu \eta^{\nu\rho}\partial_\mu \partial_\nu\partial_\rho \Big] -m\Bigg\}\Psi + H.C. \nonumber.
\end{eqnarray}
It is worthwhile to note the presence of nonlocal operators as in VSR \cite{vsr} and in Bogoslovksky spacetime.

Expanding the nonlocal operators we can rewrite Finsler-Dirac Lagrangian \eqref{randersfermionlagrangian} as
\begin{equation}
\mathcal{L}^F_{\Psi}=\bar{\Psi}(i \gamma^\mu \partial_\mu - m + \hat{\mathcal{Q}})\Psi,
\end{equation}
where $\hat{\mathcal{Q}}$ is a Lorentz violating operator. $\hat{\mathcal{Q}}$ can be written as
\begin{equation}
\hat{\mathcal{Q}}=\zeta \Big[\mathcal{Q}^{(4)} + \hat{a}^{(5)\mu}\gamma_\mu\Big] + \zeta^2 \Big[\hat{e}^{(6)}+\hat{c}^{(6)\mu}\Big] + \zeta^3 \hat{a}^{(7)\mu} + \cdots,
\end{equation}
where $\mathcal{Q}^{(4)}$ is a mass dimension four term $\mathcal{Q}^{(4)}:=-\frac{5}{2}m b^\mu \partial_\mu$,
$\hat{a}^{(5)\mu}$ and $\hat{a}^{(7)\mu}$ are CPT-Odd vector operators of mass dimension five and seven, respectively given by
\begin{equation}
\hat{a}^{(5)\mu}:=(b^\nu \partial^\mu \partial_\nu - \eta^{\nu\rho}b^\mu \partial_\nu \partial_\rho)
\end{equation}
\begin{eqnarray}
\hat{a}^{(7)\mu}:&=& \zeta^3\Big[\frac{5}{4}b^\nu b^\rho b^\sigma \partial^\mu \partial_\nu\partial_\rho\partial_\sigma - b_\mu \eta^{\nu\rho}b^\sigma b^\lambda \partial_\nu\partial_\rho\partial_\sigma\partial_\lambda\Big] ,
\end{eqnarray}
$\hat{c}^{(6)\mu}$ is CPT-even mass dimension six vector given by
\begin{equation}
\hat{c}^{(6)\mu}:=\Big[\frac{1}{2}\partial^\mu b^\nu b^\rho \partial_\nu \partial_\rho - \frac{1}{2} b^\mu \eta^{\nu\rho}b^\sigma \partial_\nu \partial_\rho\partial_\sigma + b^2 \partial^\mu \eta^{\nu\rho}\partial_\nu \partial_\rho\Big],
\end{equation}
and $\hat{e}^{(6)}$ is scalar CPT-Odd of mass dimension six defined as
\begin{eqnarray}
\hat{e}^{(6)}&:=&-  \frac{15m}{4} b^\mu b^\nu b^\rho \partial_\mu \partial_\nu   \partial_\rho.
\end{eqnarray}
Therefore, the Finsler-Dirac action in Eq.\eqref{finslerdiracaction} yields to Lorentz violating terms similar to those
of the nonminimal SME \cite{fermionsmenonminimal}. The structure of the Finsler-Dirac Lagrangian \eqref{randersfermionlagrangian} suggests the only
nonminimal SME terms arising in the Randers spacetime are the scalars and vectors.

\subsection{Gravitational field}

For the gravitational field, we adopt not only the Finsler metric $g^{F\mu\nu}(x,\nabla)$ and $\sqrt{-g^F (x,\nabla)}$ as
differential operators but the horizontal Finsler Riemann tensor $R^F_{\mu\nu\rho\sigma}(x,\nabla)$ as well.
We propose a Finslerian extension of the Einstein-Hilbert action by
\begin{equation}
\label{einsteinfinsleraction}
S^F_g:=\frac{1}{16\pi G}\int d^4x \Big\{\sqrt{-g^F (x,\nabla)}g^{F\mu\rho}(x,\nabla)g^{F\nu\sigma}(x,\nabla)R^F_{\mu\nu\rho\sigma}(x,\nabla)\Big\}.
\end{equation}
The horizontal Riemann tensor is defined as $R^F_{\mu\nu\rho\sigma}:=g^F_{\mu\lambda}R^{F\lambda}_{\nu\rho\sigma}$ where
\begin{equation}
R^{F\lambda}_{\nu\rho\sigma}(x,P):=\delta_\rho \Gamma^{F\lambda}_{\mu\nu}-\delta_\mu \Gamma^{F\lambda}_{\rho\nu} + \Gamma^{F\lambda}_{\sigma\rho}\Gamma^{F\sigma}_{\mu\nu} - \Gamma^{F\lambda}_{\sigma\nu}\Gamma^{F\sigma}_{\mu\rho},
\end{equation}
and $\delta_\mu:=\partial_\mu - N^{\rho}_{\mu}F(x,P)\frac{\partial}{\partial P^{\rho}}$ is the horizontal derivative \cite{chern}.
Considering the horizontal metric compatible Cartan connection in Eq.\eqref{cartanconnection},
the leading order connection coefficients can be written as $\Gamma^{F\rho}_{\mu\nu}(x,P)=\Gamma^{\rho}_{\mu\nu}(x)+\zeta^2 \hat{\gamma}^{\rho}_{\mu\nu}(x,P)$, where $\Gamma^{\rho}_{\mu\nu}(x)$ is the connection compatible with the background metric $g_{\mu\nu}(x)$ and
\begin{equation}
\hat{\gamma}^{\rho}_{\mu\nu}(x,P)=\left(\frac{\alpha}{F}\right)\Big[\left(\frac{\alpha}{F}\right)(\zeta b\cdot \nabla + b^2)P^\rho P^\lambda -(P^\rho b^\lambda + P^\lambda b^\rho)\Big]\Gamma_{\lambda\mu\nu}.
\end{equation}
By means of the relation $P^\mu \rightarrow \nabla^\mu$, the connection turn into the differential operator $\Gamma^{F\rho}_{\mu\nu\rho}(x,P)\rightarrow \Gamma^{F\rho}_{\mu\nu\rho}(x,\nabla)$, and hence
\begin{equation}
R^F_{\mu\nu\rho\sigma}(x,\nabla)=R_{\mu\nu\rho\sigma}(x)+\zeta^2 \hat{R}_{\mu\nu\rho\sigma}(x,\nabla),
\end{equation}
where $\hat{R}^{\lambda}_{\nu\rho\sigma}(x,\nabla):=\partial_\rho \hat{\gamma}^{\lambda}_{\mu\nu}(x,\nabla)-\partial_\mu \hat{\gamma}^{\lambda}_{\rho\nu}(x,\nabla) + \hat{\gamma}^{\lambda}_{\sigma\rho}(x,\nabla)\gamma^{F\sigma}_{\mu\nu}(x,\nabla) - \hat{\gamma}^{\rho}_{\sigma\nu}(x,\nabla)\gamma^{\sigma}_{\mu\rho}(x,\nabla)$. Substituting these operators into the Finslerian Einstein-Hilbert action, we obtain the leading order terms
\begin{eqnarray}
\label{randerseinsteinaction}
S^F_g [g] &\approx & \frac{1}{16\pi G}\int d^4x \sqrt{-g} \Bigg\{(1+\zeta b\cdot \nabla)^{\frac{1}{2}}R(x) -\frac{2\zeta^2}{(1+\zeta b\cdot \nabla)^{\frac{1}{2}}}\nabla^\nu b^\sigma R_{\nu\sigma}(x)\\
 &+& \frac{2\zeta^2 (\zeta \beta + b^2)}{(1+\zeta b\cdot \nabla)^{\frac{3}{2}}}\nabla^\nu \nabla^\sigma R_{\nu\sigma}(x) +\frac{\zeta^4 (\nabla^\mu b^\rho + \nabla^\rho b^\mu)}{(1+\zeta b\cdot \nabla)^{\frac{3}{2}}}(\nabla^\nu b^\sigma + \nabla^\sigma b^\rho)R_{\mu\nu\rho\sigma}\nonumber\\
 &-& \frac{\zeta^4 (\zeta \beta + b^2)}{(1+\zeta b\cdot \nabla)^{\frac{3}{2}}}(\nabla^\mu b^\rho + \nabla^\rho b^\mu)\nabla ^\nu \nabla^\sigma R_{\mu\nu\rho\sigma} + \zeta^2 \Big[(1+\zeta b\cdot \nabla)^{\frac{1}{2}}\hat{R}(x)\nonumber\\
&-&\frac{2\zeta^2}{(1+\zeta b\cdot \nabla)^{\frac{1}{2}}}\nabla^\nu b^\sigma \hat{R}_{\nu\sigma}(x) + \frac{2\zeta^2 (\zeta \beta + b^2)}{(1+\zeta b\cdot \nabla)^{\frac{3}{2}}}\nabla^\nu \nabla^\sigma \hat{R}_{\nu\sigma}(x)\nonumber\\
& + &	 \frac{\zeta^4}{(1+\zeta b\cdot \nabla)^{\frac{3}{2}}}(\nabla^\mu b^\rho + \nabla^\rho b^\mu)(\nabla^\nu b^\sigma + \nabla^{\sigma} b^\rho)\hat{R}_{\mu\nu\rho\sigma}\nonumber\\
 &-& \frac{\zeta^4 (\zeta \beta + b^2)}{(1+\zeta b\cdot \nabla)^{\frac{3}{2}}}(\nabla^\mu b^\rho + \nabla^\rho b^\mu)\nabla ^\nu \nabla^\sigma \hat{R}_{\mu\nu\rho\sigma}\Big]\Bigg\}\nonumber,
\end{eqnarray}
where $R(x):=g^{\nu\sigma}(x)R^{\rho}_{\nu\rho\sigma}$ and $R_{\mu\nu}(x):=R^{\rho}_{\nu\rho\sigma}$ are the background Ricci scalar and tensor respectively. Therefore, the coupling performed in the action \eqref{einsteinfinsleraction} yields to a higher-derivative theory of the background metric $g_{\mu\nu}(x)$, as in high derivative theories of gravity \cite{higherderivatives}, coupled with the background vector $b_\mu = \zeta a_\mu$.

Considering the higher derivative term on the Ricci scalar and tensor we obtain
\begin{eqnarray}
\label{finslergravitylagrangian}
\mathcal{L}^F_g [g] &\approx& \sqrt{-g} \Big\{(1+\zeta b\cdot \nabla)^{\frac{1}{2}}R(x) -\frac{2\zeta^2}{(1+\zeta b\cdot \nabla)^{\frac{1}{2}}}\nabla^\nu b^\sigma R_{\nu\sigma}(x)\\
 &+& \frac{2\zeta^2 (\zeta \beta + b^2)}{(1+\zeta b\cdot \nabla)^{\frac{3}{2}}}\nabla^\nu \nabla^\sigma R_{\nu\sigma}(x) +\frac{\zeta^4 (\nabla^\mu b^\rho + \nabla^\rho b^\mu)}{(1+\zeta b\cdot \nabla)^{\frac{3}{2}}}(\nabla^\nu b^\sigma + \nabla^\sigma b^\rho)R_{\mu\nu\rho\sigma}\nonumber\\
 &-& \frac{\zeta^4 (\zeta \beta + b^2)}{(1+\zeta b\cdot \nabla)^{\frac{3}{2}}}(\nabla^\mu b^\rho + \nabla^\rho b^\mu)\nabla ^\nu \nabla^\sigma R_{\mu\nu\rho\sigma}\Big\}\nonumber.
\end{eqnarray}
Expanding the nonlocal operators, we can rewrite the Finslerian gravity Lagrangian as
\begin{equation}
\mathcal{L}^F_g=\mathcal{L}_{EH} + \mathcal{L}_{LV},
\end{equation}
where $\mathcal{L}_{EH}:=\frac{1}{16\pi G}\sqrt{-g}R(x)$ is the usual Einstein-Hilbert action and the Lorentz violating Lagrangian $\mathcal{L}_{LV}$ are
\begin{equation}
\mathcal{L}_{LV}=\zeta \mathcal{L}^1_{LV} + \zeta^2 \mathcal{L}^2_{LV} + \zeta^3 \mathcal{L}^3_{LV} + \cdots.
\end{equation}
At first-order in $\zeta$, we obtain the Local Lorentz violating term
\begin{equation}
\mathcal{L}^1_{LV}=(k^{(5)})_{\mu\nu\rho\sigma\lambda}\nabla^{\lambda}R^{\mu\nu\rho\sigma},
\end{equation}
where the dimension five operator is defined as
\begin{equation}
(k^{(5)})_{\mu\nu\rho\sigma\lambda}:=\frac{1}{32\pi G}\sqrt{-g}g_{\mu\rho}g_{\nu\sigma}b_\lambda.
\end{equation}
At second-order in $\zeta$ we have the mass dimension four and six operators
\begin{equation}
\mathcal{L}^2_{LV}=(k^{(4)}_{\mu\nu\rho\sigma})R^{\mu\nu\rho\sigma} + (k^{(6)}_{\mu\nu\rho\sigma\lambda\xi})\nabla^\lambda \nabla^\xi R^{\mu\nu\rho\sigma},
\end{equation}
where
\begin{equation}
(k^{(4)}_{\mu\nu\rho\sigma}):=-\frac{1}{8\pi G}\sqrt{-g}g_{\mu\rho}\nabla_\nu b_\sigma,
\end{equation}
\begin{equation}
(k^{(6)}_{\mu\nu\rho\sigma\lambda\xi}):=-\frac{1}{64\pi G}\sqrt{-g}g_{\mu\rho}g_{\nu\sigma}b_\lambda b_\xi \nabla^\lambda \nabla^\xi.
\end{equation}
It is worthwhile to say that the mass dimension four LV operator $(k^{(4)}_{\mu\nu\rho\sigma})$ vanishes for a covariantly constant background vector. The dimension four local Lorentz violating operator appears in the gravitational sector of the SME \cite{kosteleckysmegravity}, whereas the dimension five, six and seven operators arise in the nonminimal SME gravity sector \cite{gravitynomminimalsme}.


From the Finslerian Lagrangian Eq.\eqref{finslergravitylagrangian} we obtain the modified Einstein equations up to second-order in $\zeta$
\begin{equation}
R_{\mu\nu} - \frac{R}{2}g_{\mu\nu} + \Lambda^F (x,b)g_{\mu\nu}\approx 0,
\end{equation}
where the anisotropic term has the form
\begin{eqnarray}
\Lambda^F (x,b)&:=& \zeta \frac{1}{8} b^{\rho}\nabla_\rho R +\zeta^2 \Big[\frac{1}{8}b^\mu b^\nu \nabla_\mu \nabla_\nu R + \nabla^\mu b^\nu R_{\mu\nu} \Big]
\end{eqnarray}
Therefore, the interaction between the background metric $g_{\mu\nu}$ with the background vector $b_\mu$ yields to a anisotropic pressure term $\Lambda^F (x,b)$.

\section{Final remarks and Perspectives}
\label{section6}

In this work we studied the symmetries of a local Lorentz violating Randers-Finsler spacetime. The anisotropic Randers-Finsler metric provides us
a way to include the Lorentz violation into the geometric structure of the spacetime and to define a Lorentz violating field theory.

We analysed the symmetries of the modified mass shell and obtained the observer transformations which preserve this deformed quadric. For a timelike or spacelike background vector, the deformed mass shell is a double elliptic hyperboloid whose major semi axis points into the direction of the background field. By diagonalizing the mass shell quadric, we obtain the observer Randers-Finsler tranformation. This transformation is the product of boost from the initial frame to the privileged frame, defined by the eigenvector of the mass shell, with the matrix formed with the eigenvalues and a Lorentz transformation.
These anisotropic Randers transformations resembles those found in Ref.\cite{changmdr}. For a lightlike background vector, the mass shell is not deformed, what shares a resemblance with the Very Special Relativity background vector \cite{vsr}. As a perspective, we point out the mass shell analysis for more complex Finsler spacetimes, such as the b-space \cite{Kosteleckyriemannfinsler,foster} and the bipartite spaces \cite{Kosteleckybipartite,euclides}, which exhibit quartic mass shell.

For the active particle transformations, we employed the analysis of the Killing vectors along the particles worldline. The horizontal conservation of the linear momentum and spin yields to the existence of these infinitesimal symmetries. For a constant background vector, we defined a translation and proper particle Randers-Finsler transformation which preserves the Randers-Finsler metric. We showed that the generators of the Randers-Finsler observer transformations are the same as those of the Randers-Finsler particle transformation. The background vector deforms the particle rest frame
providing an analogous shear and twist.
The Lorentz generators are modified by a term depending on the background vector and the particle mass, but not on the momentum. Remarkably, for a timelike background vector, only the boosts generators are modified. We argue that the more involved structure of the bipartite and b-space may produce momentum-dependent modified generators, as performed in Ref.\cite{girelli}.

The algebra defined by the generators is also modified. The structure coefficients for the momentum algebra are modified by the horizontal Finsler curvature. For a horizontal flat Finsler spacetime, the translation operators commutes.
Nevertheless, in order to the Randers-Finsler generators satisfy a Finslerian extension of the Lorentz algebra, the Finsler metric ought to be constant. The commutation relations obtained extend the one found in Ref.\cite{changmdr} and Ref.\cite{girelli} by using the Finsler metric as the structure coefficients, what provides the anisotropic factor $\frac{\alpha}{F}$ and the additional anisotropic terms. The Randers algebra bears a resemblance with the deformed Heisenberg-Weyl algebra in the $\kappa$-Minkwoski spacetime \cite{kalgebra}. This result shows the relation between deformed algebras and Finsler spacetimes, as the Bogoslovsky algebra be the deformed $SIM(2)$ algebra \cite{Gibbons}. A interesting perspective is the analysis of the Randers algebra for a nonconstant background vector satisfying the horizontal flat condition. In this case, the Finsler metric
derivatives term may provide more momentum-dependent modifications on the Lorentz algebra.

For a constant background vector, we obtained Finslerian extensions of the Casimir and Pauli-Lubanski operators. Although the momentum and Pauli-Lubanski operators are linear operators defined using the derivative operator and the background vector field, the Casimir and the squared Pauli-Lubanski operators exhibit a noncannonical squared root term. This nonlocal feature is the result of using the Finsler metric to square the operators. Instead, if the square is evaluated with the background metric, as in the SME-based Finsler structures \cite{Kosteleckyclassicallagrangians,russell}, the nonlinear character can be avoided.

We defined a Finsler field theory based on the Randers symmetries. Interpreting the Finsler metric as a differential operator acting on the fields, we showed that a minimal coupling of the scalar and vector fields with the Randers-Finsler metric, and the coupling of the fermion with the Randers-Finsler vielbein yields to nonlocal operators, as expected from the structure of the casimir operators. The presence of noncanonical or nonlocal terms in this coupling is similar to those found in VSR \cite{vsr}, Elko spinor \cite{elko} and in the Bogoslovsky spacetime \cite{bogoslovsky}.  The perturbative character of the Lorentz violating allow us to expand these nonlocal operators and find some minimal and nonminiamal SME terms. For the gauge vector field, the Randers background vector also produces the Carroll-Field-Jackiw term. For the gravitational field, the coupling of the Randers-Finsler metric with the Riemann curvature tensor also produces minimal and nonminiamal SME Lorentz violating terms. As perspectives we point out the analysis of bounds for the coupling proposed based on tests for the gauge \cite{gaugesmenonminimal}, fermionic \cite{fermionsmenonminimal} and gravitational fields \cite{gravitynomminimalsme}. The stability analysis of the fields, as done in Refs. \cite{stability}, is also an important future development. Moreover, the minimal coupling of the fields with the Finsler metric or vielbein in the Bipartite spacetime may provides additional nonminimal SME terms.


\section*{Acknowledgments}
The author is grateful to Alan Kosteleck\'{y} for useful discussions and to the Brazilian agency Coordena\c{c}\~ao de Aperfei\c{c}oamento de Pessoal de N\'ivel Superior (CAPES) (grant no. 99999.006822/2014-02) for the financial support.




\begin{thebibliography}{99}

\bibitem{KS}  V. A. Kosteleck\'{y} and S. Samuel, Phys. Rev. D \textbf{39},
683 (1989);  Phys. Rev. Lett. {\bf 63}, 224 (1989). Phys. Rev. Lett. {\bf 66}, 1811
(1991).

\bibitem{noncommutativegeometry}
S. M. Carroll, J. A. Harvey, V. A. Kosteleck\'{y}, C. D. Lane, and T. Okamoto, Phys. Rev. Lett. {\bf 87}, 141601 (2001).

\bibitem{horava} P. Ho\v{r}ava, Phys. Rev. D {\bf 79}, 084008 (2009).

\bibitem{lqg}
J.~Alfaro, H.~A.~Morales-Tecotl and L.~F.~Urrutia,
  Phys.\ Rev.\ Lett.\  {\bf 84}, 2318 (2000).
  Phys.\ Rev.\ D {\bf 65}, 103509 (2002).


\bibitem{vsr}
   A.~G.~Cohen and S.~L.~Glashow,
   Phys.\ Rev.\ Lett.\  {\bf 97}, 021601 (2006).
R.~V.~Maluf, J.~E.~G.~Silva, W.~T.~Cruz and C.~A.~S.~Almeida,
  Phys.\ Lett.\ B {\bf 738}, 341 (2014).

\bibitem{cptviolation}
  D.~Colladay and V. A. Kosteleck\'{y},
  Phys.\ Rev.\ D {\bf 55}, 6760 (1997).  Phys.\ Rev.\ D {\bf 58}, 116002 (1998).


\bibitem{smeapplications}
V. A. Kosteleck\'{y} and M.~Mewes,
  Phys.\ Rev.\ D {\bf 69}, 016005 (2004). V. A. Kosteleck\'{y},
  Phys.\ Rev.\ Lett.\  {\bf 80}, 1818 (1998).
R.~Casana, M.~M.~Ferreira, Jr, A.~R.~Gomes and P.~R.~D.~Pinheiro,
  Phys.\ Rev.\ D {\bf 80}, 125040 (2009).
V. A. Kosteleck\'{y} and A.~J.~Vargas,
  Phys.\ Rev.\ D {\bf 92}, no. 5, 056002 (2015).

\bibitem{rmp} V. A. Kosteleck\'{y} and N. Russell, Rev. Mod. Phys.  {\bf 83}, 11 (2011).

\bibitem{kosteleckysmegravity}
V. A. Kosteleck\'{y},
  Phys.\ Rev.\ D {\bf 69}, 105009 (2004).

\bibitem{bumblebee}
R. Bluhm and V. A. Kosteleck\'{y}, Phys. Rev. D \textbf{%
71}, 065008 (2005).C. A. Hernaski, Phys. Rev. D {\bf 90}, 124036 (2014).
  R.~V.~Maluf, J.~E.~G.~Silva and C.~A.~S.~Almeida,
  Phys.\ Lett.\ B {\bf 749}, 304 (2015).



%
 \bibitem{Jacobson:2000xp}
   T.~Jacobson and D.~Mattingly,
   Phys.\ Rev.\ D {\bf 64}, 024028 (2001).

\bibitem{kalgebra}
J.~Lukierski, A.~Nowicki and H.~Ruegg,
  Phys.\ Lett.\ B {\bf 293}, 344 (1992).


\bibitem{chern}
D. Bao, S. Chern, Z. Shen, An introduction to Riemann-Finsler geometry,
Springer, 1991.

\bibitem{rund}
H. Rund, The differential geometry of Finsler spaces, Springer (1959).

\bibitem{miron}
R. Miron, D. Hrimiuc, H. Shimada, S. Sabau, The geometry of Hamilton and Lagrange spaces, (Kluwer
Academic Publs., Dordrecht, 2001).

\bibitem{Kosteleckyclassicallagrangians}
   V. A. Kosteleck\'{y} and N.~Russell,
   Phys.\ Lett.\ B {\bf 693}, 443 (2010).

\bibitem{Kosteleckyriemannfinsler}
   V. A. Kosteleck\'{y},
   Phys.\ Lett.\ B {\bf 701}, 137 (2011).


%

\bibitem{foster}
  J.~Foster and R.~Lehnert,
  Phys.\ Lett.\ B {\bf 746}, 164 (2015).

\bibitem{Kosteleckybipartite}
    V. A. Kosteleck\'{y}, N.~Russell and R.~Tso,
   Phys.\ Lett.\ B {\bf 716}, 470 (2012).



\bibitem{euclides}
  J.~E.~G.~Silva and C.~A.~S.~Almeida,
  Phys.\ Lett.\ B {\bf 731}, 74 (2014).

\bibitem{ColladayMc}
  D.~Colladay and P.~McDonald,
  Phys.\ Rev.\ D {\bf 85}, 044042 (2012). Phys.\ Rev.\ D {\bf 92}, no. 8, 085031 (2015).

\bibitem{russell}
  N.~Russell,
  Phys.\ Rev.\ D {\bf 91}, no. 4, 045008 (2015).

\bibitem{marco}
M.~Schreck,
  Phys.\ Rev.\ D {\bf 91}, no. 10, 105001 (2015).
  Eur.\ Phys.\ J.\ C {\bf 75}, no. 5, 187 (2015). 
  Phys.\ Rev.\ D {\bf 92}, no. 12, 125032 (2015).



\bibitem{dsrfinsler}
   S.~Mignemi,
   Phys.\ Rev.\ D {\bf 76}, 047702 (2007).  G.~Amelino-Camelia, L.~Barcaroli, G.~Gubitosi, S.~Liberati and N.~Loret,
  Phys.\ Rev.\ D {\bf 90}, no. 12, 125030 (2014).

\bibitem{Gibbons}
   G.~W.~Gibbons, J.~Gomis and C.~N.~Pope,
   Phys.\ Rev.\ D {\bf 76}, 081701 (2007). A.~P.~Kouretsis, M.~Stathakopoulos and P.~C.~Stavrinos,
  Phys.\ Rev.\ D {\bf 79}, 104011 (2009).

\bibitem{bogoslovsky}
G.~Y.~Bogoslovsky and H.~F.~Goenner,
  Gen.\ Rel.\ Grav.\  {\bf 31}, 1565 (1999).
   G.~Y.~Bogoslovsky and H.~F.~Goenner,
   Phys.\ Lett.\ A {\bf 323}, 40 (2004).








\bibitem{randers}
 G. Randers,
 Phys. Rev. {\bf 59}, 195 (1941).





\bibitem{beem}
J. K. Beem, Canad. Jour. Math. {\bf 22}, 1035 (1970).

\bibitem{bejancu}
A. Bejancu and H. Farran, Geometry of pseudo-Finsler submanifolds. Kluwer Acadamic Publishers (2000).

\bibitem{randerscosmology}
  Z.~Chang and X.~Li,
  Phys.\ Lett.\ B {\bf 676}, 173 (2009). S.~Basilakos, A.~P.~Kouretsis, E.~N.~Saridakis and P.~Stavrinos,
  Phys.\ Rev.\ D {\bf 88}, 123510 (2013).


\bibitem{randersastrophysics}
 Z.~Chang and X.~Li,
  Phys.\ Lett.\ B {\bf 668}, 453 (2008). X.~Li and Z.~Chang,
  Phys.\ Lett.\ B {\bf 692}, 1 (2010).

\bibitem{girelli}
  F.~Girelli, S.~Liberati and L.~Sindoni,
  Phys.\ Rev.\ D {\bf 75}, 064015 (2007).


\bibitem{changmdr}
   Z.~Chang and X.~Li,
   Phys.\ Lett.\ B {\bf 663}, 103 (2008).



\bibitem{vacaru}
   S.~I.~Vacaru,
   Class.\ Quant.\ Grav.\  {\bf 28}, 215001 (2011).
  JHEP {\bf 9809}, 011 (1998).




\bibitem{pfeifer}
   C.~Pfeifer and M.~N.~R.~Wohlfarth,
   Phys.\ Rev.\ D {\bf 85}, 064009 (2012).  Phys.\ Rev.\ D {\bf 84}, 044039 (2011).

\bibitem{Perlick}
  V.~Perlick,
  Gen.\ Rel.\ Grav.\  {\bf 38}, 365 (2006).

\bibitem{Lammerzahl}
  C.~L\"{a}mmerzahl, V.~Perlick and W.~Hasse,
  Phys.\ Rev.\ D {\bf 86}, 104042 (2012). Y.~Itin, C.~L\"{a}mmerzahl and V.~Perlick,
  Phys.\ Rev.\ D {\bf 90}, no. 12, 124057 (2014).

\bibitem{gaugesmenonminimal}
  V.~A.~Kosteleck\'{y} and M.~Mewes,
  Phys.\ Rev.\ D {\bf 80}, 015020 (2009).  C.~M.~Reyes,
  Phys.\ Rev.\ D {\bf 82}, 125036 (2010). M.~Schreck,
  Phys.\ Rev.\ D {\bf 89}, no. 10, 105019 (2014).

\bibitem{fermionsmenonminimal}
   A.~Kosteleck\'{y} and M.~Mewes,
  Phys.\ Rev.\ D {\bf 88}, no. 9, 096006 (2013). A.~Kostelecky and M.~Mewes,
  Phys.\ Rev.\ D {\bf 85}, 096005 (2012).
  M.~Schreck,
  Phys.\ Rev.\ D {\bf 90}, no. 8, 085025 (2014).

\bibitem{gravitynomminimalsme}
Q.~G.~Bailey, A.~Kosteleck\'{y} and R.~Xu,
  Phys.\ Rev.\ D {\bf 91}, no. 2, 022006 (2015).

\bibitem{Carroll}
  S.~M.~Carroll, G.~B.~Field and R.~Jackiw,
  Phys.\ Rev.\ D {\bf 41}, 1231 (1990).






\bibitem{elko}
 D.~V.~Ahluwalia and S.~P.~Horvath,
  JHEP {\bf 1011}, 078 (2010).
 A.~E.~Bernardini and R.~da Rocha,
  Phys.\ Lett.\ B {\bf 717}, 238 (2012).

\bibitem{noncanonicalkinetic}
 E.~Babichev, V.~Mukhanov and A.~Vikman,
  JHEP {\bf 0802}, 101 (2008).

\bibitem{higherderivatives}
 A.~De Felice and S.~Tsujikawa,
  Living Rev.\ Rel.\  {\bf 13}, 3 (2010).
  B.~Chen and Q.~G.~Huang,
  Phys.\ Lett.\ B {\bf 683}, 108 (2010).
 C.~A.~Hernaski and H.~Belich,
  Phys.\ Rev.\ D {\bf 89}, no. 10, 104027 (2014).

\bibitem{stability}
 V. A. Kosteleck\'{y} and R.~Lehnert,
  Phys.\ Rev.\ D {\bf 63}, 065008 (2001). S.~M.~Carroll, T.~R.~Dulaney, M.~I.~Gresham and H.~Tam,
  Phys.\ Rev.\ D {\bf 79}, 065011 (2009).


\end{thebibliography}
\end{document}